\documentstyle[sprocl]{article}

\begin{document}

\title
{CONCLUDING TALK AT THE WORKSHOP ``QCD  AT THE THRESHOLD OF THE
FOURTH DECADE/IOFFEFEST"\\
\vspace{2mm}
\large\bf
  Landau's Theoretical Minimum, Landau's Seminar,
ITEP in the Beginning of the 1950's}

\author{BORIS L. IOFFE}

\address{Theoretical Division\\
Institute of Theoretical and
Experimental Physics,\\
B. Cheremushkinskaya 25, 117259 Moscow, Russia}
\date{}

\maketitle

\abstracts{In this talk I would like to share with you recollections
which refer mostly to the beginning of my professional career.
They carry an imprint of the epoch long gone...}

\section{Landau's theoretical minimum}

I start with the story of how I became   Landau's student. I was a third
year student at the Physics Department of   Moscow University.
My desire   was to be enlisted in the theoretical group, and I managed
to do so. The
professors of the Physics Department were strong in the Marxist philosophy,
but almost all   were weak in physics. Especially bad was the
situation in the theoretical group: all high class theoreticians
--- Landau, Tamm, Leontovich and others --- had been expelled. I was
dissatisfied by the quality of education at the  Physics Department, but was
in doubts, whether my abilities were enough to become Landau's student.
Finally, I collected all my courage and in the   summer of 1947 I made a decisive
step: I called Landau and asked him whether I could become his student. He
invited me  to come the next day. This was an entrance examination on
mathematics, which I  passed easily. Landau gave me the programs of
eight courses on theoretical physics. Besides, there was one more
examination on mathematics --- complex variables, special functions,
the Laplace transformation, etc. By that time only a few books of the
famous Landau
course were  published: Mechanics, Classical Field Theory, The Theory of
Continuous Media and the first (classical) part of Statistical Physics. One had
to study all other courses   by reading various
books and original papers. For example, let me display the list of books/papers
which we were supposed to  study  for the Quantum Mechanics course:

\vspace{3mm}

\centerline{\bf Quantum Mechanics}

\vspace{2mm}
\noindent
1.~Blokhintsev, {\em Introduction to Quantum Mechanics} (in Russian), Chs.
3--14, 17--22, 24;\\
\noindent
2. Kronig, {\em Striped Spectra and Molecular Structure} (in Russian);\\
\noindent
3. Rosenthal and Murphy, {\em Rev. Mod. Phys.} {\bf 8}, 317 (1936);\\
\noindent
4. Bethe, {\em Ann. der Phys.} {\bf 3}, 133 (1929);\\
\noindent
5. Pauli,  {\em Hdb. der  Phys.} XXIV-2, II, 2, 12;\\
\noindent
6. Brillouin, {\em Quantum Statistics} (in Russian) \S 124;\\
\noindent
\noindent
8. Bethe, {\em Ann. der Phys.} {\bf 5}, 325 (1930);\\
\noindent
9. Mott and Massey, {\em The Theory of Atomic Collisions} (in Russian), Chs.
2 and 5;\\
\noindent
10. Landau, {\em  Sov. Phys.} {\bf 1}, 68 (1932), {\bf 2}, 46 (1932);\\
\noindent
11. Bethe  and Peierls, {\em  Proc. Roy. Soc.} {\bf A 148}, 146 (1935);\\
\noindent
12. Breit and Wigner, {\em  Phys. Rev.} {\bf 49}, 519 (1936).

\vspace{2mm}


The papers were either  in English or  German and some of them were very long
(e.g., each of Bethe's papers was about 100 pages). Therefore, it was
implicitly assumed, that the student-to-be knew well both foreign languages,
which was very unusual at that time.

The examination proceeded as follows. An aspiring student would call Landau
and say: I would like to pass an  exam on such and such course (the order
was more or less arbitrary).

--- OK, please come on a certain day
and time.

When   students   arrived at Landau's apartment, Landau would
ask  them to leave all their books, notes, etc. in the garderobe and
invite them into a small room with a round table, with  a few
pages of blank paper on it,  and nothing else. Then Landau would
formulate a problem and leave, but every 15 to 20 minutes
he would reappear
  and look   over one's shoulder.

If he was silent, then this was a
good sign, but sometimes he would say ``hmm''  --- this was a bad sign.
I have no failed examination experience of my own. However, once, when I
was passing statistical physics, I started  solving a problem in a way that
 Landau did not expect. Landau came, looked and  said: ``hmm.'' Then he
left. In 20 minutes he came back, looked again and said ``hmm'' in  an
even more dissatisfied tone. At that moment Evgeny
Lifshitz appeared, who
also looked at my notes and shouted: ``Dau, do not waste time, throw
him out!" But Dau replied: ``Let us give him another 20  minutes."
During this time I   got the answer and it was correct! Dau
looked at the answer,   looked again at my calculations and
agreed, that I was right. After that, he and Lifshitz asked a few easy
questions, and the exam was over.

The problems given by Landau in the examinations
 were sometimes very
complicated and the student had to solve each of them in about an hour. (As
a rule, they would get  2 or 3 problems in the examination session). So one
had to have a lot of practice in advance. In order to get experience, I tried to
find problems  wherever I could. I asked Abrikosov, who had passed Landau's
minimum before me, what problems he got (but not their solutions!) and
solved them. After a few examination sessions I realized that Landau had
only a limited number of problems --- sometimes he would  give me the
same problem which he had given to
Abrikosov. I gathered that Landau understood
that his students would inform each other as to what problems had been  given, but
that did not worry him:   to estimate the student's ability  it was enough
for him to observe the process of the solution.
Here I will give you an example of
Landau's problems -- the one from  macroscopic electrodynamics. A
dielectric sphere with the electric and magnetic susceptibilities
$\varepsilon
_1$  and $
\mu_1$ is rotating with angular frequency $\omega$ in a constant electric
field ${\vec E}$ in a medium, characterized by
the parameters  $\varepsilon_2$  and $\mu_2$. The
angle between the rotation axis and the direction of ${\vec  E}$ is $\alpha$.
Find the electric and magnetic fields inside the sphere and in the medium.

It took me almost two years to  pass  Landau's minimum. (During the same
two years, I did two scientific works under Pomeranchuk's supervision). In
June 1949, after the last examination, Landau officially recognized me as his
scholar and included my name in  the list.

 \section{  Landau's seminar}

To be   Landau's  disciple   implied no
privileges,  only   obligations. That's because anybody could have
scientific discussions with Landau and get his advice. Moreover, only
a few among those who passed Landau's minimum  became his
graduate students
(I did not).  Landau's students
enjoyed  full rights as  participants of
Landau's seminar. But, again, anyone could participate in his
seminar, ask questions  and make remarks.
The obligations of the
``full-right" participants were to prepare, in a regular way, in
alphabetical order, review talks for the seminar. After each seminar
Landau would take a recent issue of Physical Review (at that time it
was not divided into sections) and point out to a speaker-to-be
which papers he was supposed to  report on  at the seminar. As a rule,
he would choose
a dozen such papers from all branches of physics. Mostly, they
were experimental or  part theoretical-part experimental. Sometimes,
it could  also be short theoretical papers, such as Letters to the
Editor, etc. The speaker  not only had to review the paper, i.e.
present its basic idea and final results, but was supposed to
understand well
how the  results were obtained, present and explain to
the audience all
necessary formulae,  including  experimental techniques, and have his
own opinion,  as to whether or not the results were reliable. In short,
the speaker was  almost as much  responsible for the reported
paper (and for errors in it!) as if he were the author.  As I
have already mentioned, the subjects of these papers were quite varied
--- from particle and nuclear physics to properties of metals and
liquids.
Landau's special love was the properties of alums.
Landau knew well all subjects (despite the fact that he almost did not read
papers, only listened  to their presentations) and put questions which
had to  be immediately and definitely answered --- general words or
statements like  ``the author  claims ...''  were not accepted.
In the audience there were always specialists on any subject,
and they also put questions and made remarks. Therefore, it was
 a hard task to present
such a talk. (Luckily, this would happen once or twice   a year).
Sometimes, when Landau  was dissatisfied with the presentation of a
paper,  he would stop  the speaker and ask him/her to go to the next issue.
If such an
event occurred two or three times during a given  report, Landau
would say: ``You
did not prepare your lesson! Who is the next speaker?''

In the worst
cases, when the speaker failed a few times, he was ostracized ---
excluded from the list of the seminar participants, and Landau
would  refuse
to have discussions with him, but, of course, he (the ostracized
person) could attend
seminars. (I remember two such cases --- in one case the speaker
was a famous physicist, V.G.~Levich,
who eventually became  a Member of the
Academy of Sciences). Only after a long time, a year or more,
and after being advocated by the most respected seminar participants,
could such a person  be pardoned by Landau.

The presentation of a theoretical report would proceed  differently. A person,
who wanted  to present a theoretical investigation at the seminar (his own
or from the literature) was first supposed   to tell the story
to Landau privately.
If Landau agreed with the basic points of the work, then the talk at the seminar
would be allowed. During the talk, Landau gave clarifying comments and quite
often his explanation  of the work was strongly different from that of the
author. A hot discussion would then often follow. One could  hear  from Landau:
``The
author, in fact, did not understand what he did." Landau's understanding in
all cases was quite original and for normal people it was not easy to follow his
line of reasoning. For me (and not only for me) it would require  a few hours
(sometimes, a few days) before I could understand how deep his remarks were,
which often would turn  the problem upside down and shed light  on it from a
different side. Theoretical talks  freed a speaker from obligatory presentations
of Physical Review papers; therefore, it was a serious stimulus to present
theoretical talks at Landau seminars. (Pomeranchuk, for example, never did
reviews, since he always presented theoretical talks). Sometimes, speakers
from the outside, who were not from Landau's school, presented theoretical talks. Up
until 1955 no foreign physicists visited Moscow. So, by outside theoreticians, I mean
the ones from FIAN, Mathematical Institute and Moscow University
(Bogolyubov, Gelfand), as well as from Leningrad and Kharkov.

This was  the normal routine of Landau's seminar at  the end of the 1940's
and  in the beginning of the 1950's. There were exceptional
persons, however: Ginzburg and Migdal. Once upon a time Landau said about
Ginzburg: ``Ginzburg is not my disciple --- he just jumped onto the
bandwagon." Indeed, Ginzburg came from   Tamm's school, but was a very
active participant of Landau's seminar. He did not follow the standard
seminar routine with presentations of review talks, etc. Each
time  he arrived, he was full of new facts and ideas and
presented them with brilliance and sharp wit. I vividly remember
his impressive talk on supernovae,  with a historical introduction
on their observations in ancient Babylonia, Egypt and China. It
was no accident that the famous phenomenological theory of
superconductivity, the predecessor of many modern models of
spontaneous symmetry breaking,  was done by Ginzburg and Landau.

Another exceptional person was Migdal. His name is absent in the
list of Landau's
disciples, written by Landau himself: he did not
pass  Landau's minimum, but he was a full-fledged participant of
the seminar. It was only Migdal whom Landau allowed to be late to
the seminar and, nevertheless, enter the hall through the front
door. As a rule, the seminars would  start  just in time,
up to a minute. But sometimes Landau would say: ``Let us wait
for  five
minutes --- these are Migdal's five minutes." One day, in the middle of
the seminar, the front door of the hall opened and a person wearing a
 fireman's helmet and jacket appeared. ``Get out!  Leave the hall
--- we will perform here anti-fire exercises!''  ---  exclaimed the man in
a severe
tone of voice.  Lifshitz jumped up and shouted: ``We have a seminar
here every Thursday! You have no right!''

``Get out!'' -- repeated
the man inexorably. People started   standing up and moving to the
doors. Then the man took off his helmet and the thread, which held
his nose up --- it was Migdal!

Another good joke was a letter from Pauli, which Landau had received
through Pontecorvo. It was in 1958. At that time Landau was enthusiastic
about  Heisenberg's recent papers, where a universal nonlinear fermion theory
was suggested. In a short letter, which Landau read at the seminar, Pauli
claimed that he had found new arguments in favor of Heisenberg's theory
and,  moreover , there were  new experimental facts supporting it. The facts,
however, were not mentioned explicitly, and there was only a hint about their
origin. The majority of the seminar participants became very excited.
Somebody even went to the blackboard and tried to imagine which
experiments they could be. Meanwhile, Migdal took the letter, read it carefully
and said:  ``Please, look. If you read the first letters in each line, you
will find the
Russian word  ``duraki" (fools);  what could that mean?"

In 1950--1951 the first experimental data on the pion production in
$pp$ collisions appeared. Because of the low energies available,
the data referred only to the threshold region. Migdal had
immediately formulated the theory of this phenomena: he had
demonstrated that $pn$ interaction in the final state dominated
here and proved that this interaction reduced to $pn$ scattering phase
in the $S$-wave. He had also calculated the ratio $\sigma(pp \to
pn \pi^+)/\sigma(pp \to p d \pi^+)$, which was in  good agreement with
the data. Migdal   presented a talk about his results at
Landau's seminar, which was met with great enthusiasm. However, he had failed
to publish  this paper. In Kurchatov Institute (Laboratory No.2), where
he worked, the paper was classified and its publication   forbidden.
In the US the same results were obtained by K.M. Watson (Phys. Rev.
{\bf 88} (1952) 1163) a year later and were  called ``the Watson effect."
Migdal managed  to publish his paper only in 1955
(JETP {\bf 28} (1955) 10). In ITEP, the papers on related
subjects,  $\pi^-$ capture in hydrogen and deuterium,
photoproduction of pions on deuterium, etc.,  were not classified and were
published thanks to Alikhanov.

The third exceptional person was Pomeranchuk. I will dwell on
him later.

\section{My senior thesis under
Pomeranchuk's supervision}

 Now I return to 1947. I was a student in the
theoretical group for a very short time ---  about a month. Then an
order from the Dean's Office came,  and I and a few of my friends were
transferred to a Department called  ``The Structure of Matter." This name
was a camouflage: in fact, this meant nuclear physics. My friends
(David Kirzhnitz among them) and I did not want  to go into this
field. For a month we tried to resist:  we visited the Dean
of the Physics Department,  a few times,   arguing in various ways.  But the
order was strong, and we had to submit to   force. Only later, after a year or
two, I realized that, in fact, this was my luck. Just because of this
transfer, I became what I am now: if I had  stayed in the
theoretical physics group, then, probably, I would have faded away.
The Structure of Matter Chair belonged to  I.M.~Frank, the future
Nobel Prize Laureate. Although the Chair was
mostly experimental,  a theoretical course was also offered there.
Moreover, it was possible to do a theoretical diploma   and take,
as a supervisor,  anybody involved in the atomic project. (The rule of
the  Theoretical Physics  Chair was that a supervisor had to be from
this Chair). I wanted to have a supervisor from Landau's school, and by
chance, chose  Pomeranchuk.  (I did not know him
before). I called him, introduced myself, and told him, that I am a
student at the Physics Department, I am taking Landau's examinations
and have already passed three of them: mechanics, classical field
theory and math-II. Pomeranchuk invited me to come for a
conversation.  When I   came, after a short conversation,
Pomeranchuk agreed to  supervise  my diploma work. He
said, however,  that   first I had to pass all of Landau's examinations. I believe
that there were two motivations which prompted Pomeranchuk's decision.
First,  I
was a student at the Physics Department of Moscow University during  the
time when   Landau's name was notorious there, and the fact that a student
would like to have him --- Pomeranchuk --- as a supervisor, was not trivial.
And, secondly, in 1948   only a dozen persons passed Landau's
minimum, and all of them were outstanding physicists (with the
exception of the last two --- Ter-Martirosyan and Abrikosov --- who had
passed the minimum just before me and had no time to manifest
themselves). Later, Pomeranchuk told me that he was surprised by
the way I was dressed:  it was a very cold winter day of 1948, but I came
very poorly dressed. In turn, I was surprised by the absence of furniture
in   Pomeranchuk's apartment: a bed covered by a soldier-type blanket, a
table, a bookcase, and nothing else.

Thus, I was continuing my preparations for Landau's quantum mechanics
examination.

Here is an episode, characterizing the levels of education at   Moscow
University and Landau's minimum. In the spring of 1948, the time   came
to pass the quantum mechanics
examination at    Moscow University. The lecturer
was Blokhintsev, but I did not attend his lectures. I was busy
studying quantum mechanics according to Landau's program, and at that
time I estimated the level of my knowledge  of the subject to be low:
I had to work much more. One day, I   met D.~Shirkov , who was a
student in the theory group. ``I am going to pass Blokhintsev's quantum mechanics
exam. Do you want to join me?''   he asked.

 ``OK, I'll
  put Blokhintsev's book in  my bag in order to look at it, just in
case,''  I replied.

We   passed this examination successfully, I   got an
A, and  Shirkov a B. I managed to  pass  the same examination
given by  Landau only in
September.

Since I was successfully progressing with my examinations, in
the late autumn
of 1948 Pomeranchuk   formulated a problem for my senior thesis work:
it was the calculation of neutron polarization in the scattering
off nuclei due to interference of the Coulomb and nuclear
scattering. The calculation was based on  Schwinger's paper, but
 I had to invent  some elements on my own. The second part of the work
was the calculation of the neutron depolarization in the course of
the neutron deceleration in the medium. This part was merely educational:
I had to study the theory of neutron moderation   as well as
some aspects of  the theory of nuclear reactors. During this winter, until
March of 1949,  I had finished my senior thesis and almost finished
Landau's minimum (except for the last examination ---  the theory of
continuous media, which I  passed in June of 1949). So, Pomeranchuk
gave me a new problem: to calculate the cross-section of $e^+e-$
pair production on nuclei by linearly polarized $\gamma$-quanta
and the related cross section of {\em bremsstrahlung} of polarized
$\gamma$. At that time there was no Feynman diagram technique --- the
famous   papers of Feynman
were published at the end of 1949. Therefore,
I did the calculations in the old Heitler technique with the
account of the electron transitions to negative-energy states, using
non-covariant Dirac matrices, etc. You  can see how complicated the
old technique was, looking at the original Bethe-Heitler paper,
where the calculation of the electron {\em bremsstrahlung} was
performed. (The calculation  of polarized $\gamma$ {\em
bremsstrahlung}
was not easier!) To give drastically  different problems to students was
typical of Pomeranchuk's (as well as Landau's) style: a student had to be able
to solve problems in many (if not in all) fields of physics.
Pomeranchuk recommended that I write two short papers: one on
the neutron polarization and another on  $e^+e^-$ pair production
by polarized photons. I did that, but, probably, he   forgot  about
this, and I hesitated to remind him. Thus, these papers were not
published. (They are being published in this book, in the English
translation from the
 Russian original manuscripts.)

 Later, in the 1950's, a few
papers appeared  in which calculations of $e^+e^-$ pair
production by polarized photons and electron {\em bremsstrahlung} of
polarized photons were performed, using the Feynman technique. I felt
sorry
 that I did not publish these papers.

In the spring of 1949 Pomeranchuk   introduced me to A.I. Alikhanov, the
Director of Laboratory No.3 (now ITEP),  as a person whom he would like to
take to  the Theoretical Physics Division of Laboratory No.~3. Alikhanov had a
custom --- to have   a prior conversation with any
prospective new worker in the
Laboratory. After a short conversation, Alikhanov   signed a letter,
requesting  my assignment  to  Laboratory No.~3 after my
 graduation from   Moscow
University.
This was an
extraordinary case. The anti-Semitic campaign was   in full swing,
and I was the {\em only}   Jewish
student  from the whole Physics Department who
  got an appointment in Moscow, and in a good place. All others were
sent far away (for instance, my friend Kirzhnitz was sent to a factory
in Gorky), or  got  no jobs at all.

\section{  ITEP in the 1950's}

On January 1, 1950 (a symbolic date ---  the beginning of the second half of
the 20$^{\rm th}$ century!) I started my work in  the Laboratory of
Theoretical Physics of ITEP. The Head of the Laboratory was Pomeranchuk. In
the beginning, Pomeranchuk  ``lent"  me to the ITEP Vice-Director
V.~Vladimirsky. I had to calculate electric fields in the electron linear
accelerator, which he wanted to construct. I did not like this job: I had no
idea on how to calculate the electric field for complicated configurations of
electrodes, and this work seemed  to me to be very gloomy. So, instead of doing it, I
read the papers of Feynman, Schwinger and Dyson that had just  appeared
 (I   translated  some of them to   Russian,  and the translations
were published in Russian review journals).\,\footnote{It should be
mentioned that  getting American physics journals in Moscow was a problem
at that time: they would often come with   great delay,
and  sometimes they were
stamped ``classified." We knew that   they were delivered   illegally,
through Sweden.}

I wanted to educate myself in the new approach to quantum
electrodynamics (the Feynman diagram technique, renormalizations, etc.).
At
that time nobody   in Moscow was proficient in these new QED methods,  and
only a few people (Galanin, Abrikosov, Khalatnikov, maybe, somebody else) learned
them.  Such a situation ---
 neglecting  my job in favor of Feynman, Schwinger and Dyson --- could not
continue for a long time and was destined to end in a scandal. But I
was ``lucky" again. An order came from the highest level (probably,
from Beria, or, maybe, even from Stalin himself) --- the Institute
was supposed to
present in the shortest time --- within a few weeks ---  a project of
the heavy-water nuclear reactor on enriched uranium, for tritium production.
All theoreticians,   including myself, were mobilized to do the physical design
of this reactor. I was returned under Pomeranchuk's guidance, and starting
from this time (the spring of 1950) I   worked  for many years  in parallel
on elementary particle physics and on nuclear reactor  design.

\vspace{2mm}

There were three principles which Pomeranchuk put in the basis of the work of
Theoretical Laboratory:

1. ``The Directorate must be respected." This meant that all
problems formulated for theoreticians by the Institute Management
and devoted to applied physics, such as nuclear reactor design, had to
be solved with priority and    full responsibility; any errors
had to be   completely ruled out.

2. ``The experimentalists must be respected." This meant that if an
experimentalist came
with a question to our  Theoretical division,   or asked
for   help, the question had to  be answered,
and assistance provided, even
if this required a complicated calculation.

3. ``You may do science from 8 p.m. till 12 p.m." This meant that young
people, even if they were busy doing their jobs, according  to the
points 1 and
2 above, had to find time for {\em the} science (i.e. purely theoretical work).

\vspace{2mm}

In one particular aspect Pomeranchuk and I had something in common --- we
had a common hobby,  reading newspapers. At that time almost nobody read
newspapers regularly ---  there was no information there. All
newspapers  were filled with articles
that would  start  with  words like ``New
successes in the production of ... are  achieved at the factory ... "
or  ``The worker
Ivanov (Petrov, etc.) in one  shift produced ..."
In order to get information
from   newspapers  one had to be a professional in this business, and we ---
Chuk and me --- were. (From now on I will call him    Chuk, as many did.) In
the morning, just after arriving at the ITEP, Chuk would
come into  my office and
ask:

 --- `` Have you read {\em Pravda}  today?"

--- `` Yes,"    I would reply.

--- ``And what did you pay attention to?"

---  `` To  a small article on  the third page."

---  ``Oh!"     and Chuk would raise  his forefinger.

---  ``And you?"

--- `` Probably,  the same, the report on the meeting of the Voronezh
District Party Committee."

--- ``Yes."

--- ``And what was interesting for you in this report?"

--- ``The greeting to the Politbureau."

---  ``Oh!"  And Chuk's forefinger would go up up again.  ``What
concretely?"

---  ``The ordering in which the names of the Politbureau members were
listed."

We understood  each other well. From this ordering one could obtain  an
idea of who was going up or down in the Politbureau, and
estimate political trends.

In 1950 all members of the ITEP    Theoretical Physics Laboratory ---
V.~Be\-res\-tetsky, A.~Galanin, A.~Rudik and myself ---
were intensively studying new
methods in QED. Pomeranchuk strongly supported this activity, but in1950
and in the first half of 1951 he did not participate in it too much  himself: he
was busy with  other problems --- in 1950/51 he was sent for  half a
year to Arzamas-16, to work on the hydrogen bomb project. Landau was
sceptical of the new trends in QED. He did not believe  that the problems of
infinities in quantum field theory could be circumvented by
the mass and charge
renormalization. Two attempts to present Feynman's papers at Landau's
seminar failed: the speakers were thrown off the podium after 20
or 30 minutes of talking. Only the third attempt succeeded (if I remember
correctly, this was in 1951 or even in 1952). But still he had no interest in
these problems: the dominating subjects on his seminars were what we called
``alums."

Landau called me a ``snob." He repeated this even in public: ``Boris is snob!" The
meaning of his words was that I did not want to solve real physical
problems and, instead, preferred to study a refined theory. His words
had no
influence on Pomeranchuk, because we were allies, but
 --- what was the worst ---
he said the same thing to  Alikhanov, the ITEP Director.  And for Alikhanov,
Landau's evaluation of anyone in theoretical physics, had the highest
weight. So, Landau's words could have resulted in undesirable consequences for me.
Fortunately in this case Alikhanov already had his own opinion. He knew
very well that I
was performing calculations of nuclear reactors for him (as well as
calculations of   his experimental set-up) and by no means thought I was a snob.

\section{  Pomerancuk's seminar}

Pomeranchuk tried many times to convince Landau to shift his interests to
QED and mesonic theories. Once in a while,
he would repeat:

 ``Dau, there are a lot of problems here. They are hard, but
they are just for a person of your class!"

In response,  Dau would say:

``I know my abilities ---
  to solve the problems of infinities in field theory is above them."

 In fact,
contrary to  common belief, Landau was very modest in his self-evaluation.
He
 underestimated rather than overestimated his abilities and achievements.
Experimental facts in particle physics were always reported on  Landau's
seminar, but theoretical papers were  reported only as an exception.

Then Pomeranchuk
decided to organize a separate theoretical seminar devoted to quantum field
theory and particle physics. The seminars could not proceed at ITEP, since all
participants had to  have permission to enter ITEP territory, and by no means
did everyone  have such a permission. Therefore, Pomeranchuk
made an arrangement
 with
Landau that the seminar will proceed at the Institute of Physical Problems, on
the same day of the week, Thursday, as Landau's seminar, but two hours
earlier. Pomeranchuk nominated me as the seminar secretary. The first
meeting took place on  October 1, 1951. I reported on the famous
 paper of Dyson  at this meeting. Alikhanov, as  the ITEP Director, asked me
to present to  him an official letter regarding  the creation of a new seminar, and
I did this. (This document still exists). Almost all famous theorists
participated in the seminar.   The main papers on quantum
field theory were reviewed followed, as a rule, by heated debates.
Sometimes  Landau would peep   in  the hall from the door. Chuk would
invite  him: ``Please come in, we are discussing this and that". But Landau
would only condescendingly smile:
``If young people want to spend time on
nothing, then let them do it". With time   the number of participants of
Pomeranchuk's seminar was increasing, as well as   the enthusiasm and
excitement around the problems under discussion. This excitement
eventually spilled over into Landau's seminar which, as I already mentioned,
used to start just after Pomeranchuk's  seminar. Then
Landau  decided: his seminar had to precede, not follow Pomeranchuk's.
 In 1953, when the restrictions regarding the admission of outsiders to ITEP
were somewhat relaxed, the seminar was transferred to ITEP. It still exists,
to this day, every
Monday, at 3.30 p.m. (except holidays),
the doors of the main ITEP conference hall
open for the  ITEP   theoretical seminar.

\section{ The history  of the making of Landau, Abrikosov and
Khalatnikov's papers}

Aleksei Galanin and myself educated ourselves in the calculation of
  radiative corrections in QED and meson theories, and in performing the
mass and charge renormalization --- first, at lowest order, then at
higher orders. I succeeded in writing an exact infinite system of
coupled equations for the Green's functions in the meson theory. In the
paper by Galanin, Pomeranchuk and myself the mass and charge
renormalization was performed in this system of  coupled equations.
The solution of this system of equations was shown to  have no
infinities after renormalization, it had to  be finite. However,
we did not succeed in solving this system in a recurrent way by
cutting it off at some fixed number of equations. After the cut-off the
infinities reappeared;  in order to get rid of them, one
had to  sum the whole infinite series.

Upon calculating several  first-order corrections in perturbation theory
Ga\-la\-nin and myself  realized that
 large logarithms of the type
$\ln (p^2/m^2)$ appear
in  the polarization operators and
vertex function far off-shell,  $p^2\gg m^2$.
In the first order one deals with $ln(p^2/m^2)$,
 in the second order with terms proportional to
$[\ln(p^2/m^2)]^2$, in the third order   $[\ln(p^2/m^2)]^3$,
and so on. In this aspect
the paper by S.F.~Edwards (Phys. Rev. {\bf 90}  (1953) 284)
was important for us. Edwards considered the equation for the
vertex function in the ladder approximation and demonstrated that
at the $n$-th order the terms $\sim(e^2 \ln p^2/m^2)^n$ appear.

In the 1950's Landau visited ITEP every Wednesday. He participated ---
  very actively ---  in the ITEP Wednesday experimental seminars, steered by
Alikhanov. After the seminars he used to come to the
theorists office, for discussions  which would normally
go on for 1 or 2 hours.
In these discussions we  explained to Landau the
situation with the radiative corrections and he  came up with the idea to
sum up the leading logarithmic terms, i.e. the terms $\sim(e^2 \ln
p^2/m^2)^n$ in QED. Initially, when Landau formulated his idea, he
believed that he would find in QED  what is now called
  asymptotic freedom. These expectations were
formulated in the first   papers by Landau, Abrikosov and
Khalatnikov which had been  sent for publication before the final
result was obtained. At one of the following Wednesday visits Landau
showed  us their result,  confirming  his expectation: the
effective charge in QED was decreasing with energy. Galanin and myself
  decided to check their calculation, because we had a desire
to use this idea in our coupled system of the
renormalized equations.
(We did this later, in collaboration with Pomeranchuk). But the
first-loop calculation demonstrated the opposite behavior. The
effective charge was increasing with energy! Next Wednesday
we   told Landau about this and convinced him that we  were
right. Landau, Abrikosov and Khalatnikov's paper  which was already
prepared for publication, had a sign error, drastically changing the final
conclusion. S.S.~Gershtein who worked at the Institute of Physical
Problems at that time, wrote later in his memoir
that upon returning from ITEP, Landau
said:

--- ``Galanin and Ioffe  saved me from shame."

After the publication of  Landau, Abrikosov and Khalatnikov's
 papers, in approximately one year, Landau   got a letter from Pauli. In this
letter   Pauli informed him that his graduate student, Walter Thirring, had found
an example of a theory, in which there was no zero charge problem ---
the theory of the scalar meson-nucleon interaction. The manuscript of
 Thirring's paper was attached to the letter. Dau   gave  this
paper to Chuk, and Chuk   asked me to check the paper. I
studied Thirring's paper and   came to the conclusion that it was
wrong. The origin of the  mistake was that the Ward identity
arising from differentiation over the nucleon mass, was exploited, which
in fact was violated by renormalization. I   told
Chuk about this.

--- ``You should write a letter to Pauli'' --- was
Chuk's response.

I   hesitated: to write to Pauli, that his graduate student had
made a mistake and he, Pauli, had overlooked it!   Chuk insisted
and, finally, I wrote a letter ending up  with the signature:
``Respectfully yours ...'' The answer I received was not from Pauli,
but from Thirring. He  accepted his error, and  Thirring's paper was never
published.

\section{ Papers on $P$, $C$, $T$ non-conservation}

Now I would like to tell you about another episode
which adds important touches to
Landau's portrait.

 In 1955-56 the   $\theta$ -- $\tau$  puzzle   agitated all
physicists. The $K$-meson decays in  2 and 3  pions had been
observed experimentally. Under the condition of
 parity conservation, which was taken for granted
at that time, one and the same kaon could not
 decay in  2 and 3 pions simultaneously. For this reason
most  physicists believed that $\theta$ and $\tau$ were two different mesons.
As the precision of experiments grew, however, it
became clear  that their masses coincided. At that time,
in the spring of 1956, Lee and Yang came up with their revolutionary
paper in which they put forward  a parity
non-conservation hypothesis which explained the $\theta$ --
$\tau$ puzzle. Moreover, Lee and Yang calculated parity non-conservation effects in
the $\beta$ decay and in the $\pi \to \mu \to e$ cascade.

Landau vigorously rejected the
possibility  of parity non-conservation, saying ``space
cannot be asymmetric!" Pomeranchuk preferred the hypothesis of
parity-degenerate  doublets of strange particles.  A.P.~Rudik and
I decided to calculate some additional effects based on the
assumption of parity non-conservation in weak interactions,
other than those considered
by Lee and Yang. We decided to examine $\beta$ -- $\gamma$
correlations. I   made an estimate and found that the  corresponding effect
had to  be large. Rudik turned to detailed calculations. In some time
he came to me  and said:

``Look, Boris, the effect vanishes!"

``This
cannot be the case,''  I replied.

We began trying to make sense of this result.    I observed
that Rudik, being a well-educated theorist, had imposed  the condition of $C$
invariance on weak interaction Lagrangian.
 As a
result, the coupling constants in front of the
 parity-nonconserving terms turned out
to be purely imaginary. The constants in Lee and Yang's papers were
arbitrary complex numbers. (If one assumes them to be purely
imaginary, then all   parity-nonconserving effects
disappear.)

A question arose as to the connection between $C$ and
$P$ invariance. I discussed this problem with Volodya Sudakov;
an earlier
paper by Pauli surfaced in this conversation. Although I had read
this paper previously, I  had forgotten  about it. In part, the
reason was that Landau regarded this paper with
scepticism  --- he
believed that the $CPT$ theorem was a trivial relation
satisfied for any Lagrangian and, for this reason, no physical
consequences could follow from the $CPT$ theorem.
I noted that    Lee and Yang's
paper  did not mention the $ CPT$ theorem at all, and nothing was said on
the connection between $C$, $P$ and $T$ invariance.
I   read   Pauli's paper
  again, more attentively than the first time, and it became clear to me
immediately that if $P$ were violated, then either $C$ or $T$, or   both
had to be violated with certainty.

This observation gave rise to the following idea:
  two $K^0$ mesons with drastically different  lifetimes  may appear
only provided one of the invariances, $C$ or $T$,
takes place,
at least approximately. Rudik and I   considered a number of effects
and observed that $P$-odd pair correlations of spin and momentum ($\sim
\vec{\sigma}{\vec p}$) appear if    $C$ is violated and $T$ is conserved.
In the opposite case they are absent. (In my subsequent  paper I
  proved this theorem in a general form, and found  the
type of $P$-odd terms corresponding to $T$ violation.) We
wrote a paper and told its contents to L.B.~Okun. Okun made a
very useful remark that analogous effects which unambiguously
 differentiate models
with $C$ invariance from those  with $T$ invariance,
 appear   in $K^0$
decays into pions too. We  included this remark in the paper and I
suggested to  Okun to become a co-author. At  first he refused, saying that
such  a remark deserves a mention in the acknowledgments, but later I
persuaded him.

After that  I reported our results to Pomeranchuk. Pomeranchuk
  decided that we had to tell them
 to Dau --- immediately,   next Wednesday.
On Wednesday, Dau's first reaction was to refuse  to listen.

``I
do not want to hear anything about parity nonconservation. This is
nonsense!"

Chuk persuaded him:

``Dau, have patience for about 15
minutes, listen to what young people
have to say."

 With heavy heart Dau
agreed. I spoke  not for long, perhaps, for half an hour. Dau kept
silent, and then went away.  Next day in the morning Pomeranchuk
called me:
Dau   solved the parity non-conservation problem! We
were supposed to come to him immediately.

By that time both  of Landau's papers
--- on the conservation of the combined ($CP$)
 parity and on two-component
neutrinos, with all formulations, were already ready.

Our paper and that of Landau were sent for publication
prior to experiments of Wu {\em et
 al.}, where    the electron asymmetry
  in polarized nucleus decay was observed (i.e. the  correlation between
nucleus' spin and electron's momentum). In this way the parity
non-conservation was discovered. Our
results then implied  that  the $C$ parity was not conserved
in  the $\beta$ decay either. The corresponding note was added in proof  in
our paper. An analogous statement was made in the paper by Wu {\em et
 al.}, who
referred to the paper by Lee, Oheme and Yang, which, in turn,
 was published {\em after}
our paper. In their Nobel lectures Lee and Yang emphasized our
priority
 in this problem.

 Landau considered the $CP$ conservation to be the exact law of nature;
he did not admit the possibility of
 its violation. Concerning $CP$,  Landau would say exactly  the same words
on the space
asymmetry as he used to say previously with regards to
$P$  violation. I  constructed an example of the
 Lagrangian in which $CP$ was violated, and nothing  bad happened to
the vacuum, and
 tried to change  Landau's mind, but he did not want to listen.

\end{document}